\documentstyle[12pt,aasms4]{article}
\slugcomment{{\bf To appear in}, {\em Astrophysical Journal, Letters to the Editor}}
 
\begin{document}
\title{Do dwarf spheroidal galaxies contain dark matter?}
\author{A. Burkert}
\affil{Max-Planck-Institut f\"{u}r Astronomie\\
K\"{o}nigstuhl 17, D-69117 Heidelberg, Germany}
\authoremail{burkert@mpia-hd.mpg.de}

\begin{abstract}

The amount of dark matter in the four galactic dwarf spheroidals
with large mass-to-light ratios is investigated. 
Sextans has a cut-off radius which is equal to the expected tidal
radius, assuming a high mass-to-light ratio. This satellite 
very likely is dark matter dominated.
Carina, Ursa Minor and Draco, on the other hand, cannot contain a
dominating dark matter component if the observed 'extra-tidal' stars
are located exterior to the tidal radii of these systems. 
The evidence for tidal stripping in the absence of dark matter is 
also supported by the fact that the observed cut-off radii of all
three satellites are equal to their tidal radii, assuming a low, globular
cluster like mass-to-light ratio.
The large velocity dispersions of these galaxies, on the other hand,
provide strong evidence for a massive dark matter component. In this
case, the 'extra-tidal' stars lie deeply embedded in the dark matter potential wells
of the satellites. These stars then would represent a gravitationally bound,
extended stellar component with unknown origin. 
\end{abstract}

 
\section{Introduction}
Nine dwarf spheroidal galaxies (dSphs) are known to orbit the Milky
Way.  Like globular clusters, the dSphs
have low luminosities, of order $10^5 - 10^7 L_{\odot}$.
Their core radii are however typically an order of magnitude 
larger than in globular clusters, leading to very low surface brightnesses
(see the review by Gallagher \& Wyse 1994). 
As a result, the Milky Way's massive dark halo exerts a
strong tidal force on these objects, which ultimately might lead to
their disruption. Indeed, at least one of them, 
Sagittarius (Ibata et al. 1994, Mateo et al. 1995), 
is known to be currently torn apart.

Recent theoretical and observational work has focussed on the
dark matter content in dSphs. This question is 
very interesting because of two reasons. First, 
a detailed investigation of the dark halo structure in these nearest 
galaxies could provide important information on the formation of dark
matter halos and on the nature of dark matter. Second, the origin and
future fate of dSphs within the tidal field of the Galaxy
is strongly coupled with the question of their dark matter content.

The origin of the galactic dSphs is probably closely related to
the formation history of the Milky Way. Cosmological models of
hierarchical structure formation in a cold dark matter scenario
naturally produce satellite systems around massive galaxies, which
represent those galactic building blocks that manage to
escape merging and remain gravitationally bound for a Hubble time.
In this case, we would expect that dSphs, like other galaxies,
contain a dark matter halo.
Another possibility has been suggested by Lynden-Bell (1982),
who noted that the dSphs and the Magellanic Clouds are not
uniformly located on the sky, but occupy two great circles.
They therefore might represent the debris of two massive, tidally
disrupted progenitor satellites. 
As noted by Barnes \& Hernquist (1992), dwarf galaxies 
could indeed form inside tidal arms during galaxy-galaxy collisions. 
These systems however do not contain significant amounts of dark matter. 

An answer to the dark matter question might be given by
observations of the stellar velocity
dispersions in dSphs, which would provide an estimate of their virial mass.
From this, a mass-to-light ratio $M/L$ can be derived. The observational
uncertainties in the central velocity dispersions of the dSphs
are still very large. It is however clear that most 
dSphs have $M/L$ values that are much larger than in globular clusters.
The four galaxies with exceptionally high $M/L$ values and
galactocentric distances smaller than 100 kpc are Carina, Draco, Sextans
and Ursa Minor (table 1). These systems seem to contain large amounts
of dark matter within their visible radius.

This conclusion has been questioned by Kuhn \& Miller (1989;
though see Pryor 1996), who 
argued that a resonance of the tidal force with collective
oscillation modes of a dSph could significantly heat the stellar system,
resulting in large velocity dispersions and large apparent mass-to-light
ratios, even without a dark matter component.
As a result of this interaction, the system expands beyond
its tidal radius, particles are lost and the dSph eventually dissolves.
As Kuhn \& Miller (1989) note, without a dark matter component
the dSphs with high apparent
$M/L$ and large central velocity dispersions would be gravitationally
unbound even in the core regions and should disperse on a free
expansion timescale $\tau_{dyn}$, which is  of order
$10^8 - 10^9$ yrs. 
Kuhn (1993) made however the interesting point that the lifetimes of not
internally bound satellites could actually be much longer than $\tau_{dyn}$.
The details are complex and depend on the initial velocity 
distribution of the particles.  Systems with a small velocity dispersion
along the line connecting the dSph and the galactic center and with a
large dispersion tangential to this will disperse on timescales
smaller than $\tau_{dyn}$. Configurations with the reverse
dissolve more slowly than expected from free expansion.
If the timescales for disruption become long enough, 
a large fraction of the dSphs could 
actually be unbound but not yet dispersed.  Then 
the large velocity dispersions, observed in the four galactic dSphs
are no direct signature for dark matter.

Kuhn \& Miller (1989) studied the response of a stellar system to an external,
oscillating tidal field, neglecting coriolis terms.
Oh et al. (1995) published detailed numerical
computations of the tidal disruption of dSph satellites in a
realistic, logarithmic Galactic potential, taking into account coriolis forces. They
started with isotropic King profiles (King 1966) and initial cutoff radii, 
$r_k$, equal to or larger than the tidal radii at perigalacticon
(King 1962). The tidal radius for a logarithmic 
Galactic potential is (Oh et al. 1992)

\begin{equation}
r_t = a \left( \frac{M_{dSph}}{M_G} \right)^{1/3} 
    \left( \frac{(1-e)^2}{[(1+e)^2/2e] \ln [(1+e)/(1-e)] +1}
    \right)^{1/3},
\end{equation}

\noindent where $e$ and $a$ are the orbital eccentricity and the semi-major axis,
respectively. $M_{dSph}$ and $M_G$ are the mass of the dSph and of the
Galaxy inside $a$, respectively. 
In contradiction to Kuhn (1993),  Oh et al. find that the unbound
but not yet dispersed systems have velocity dispersions which 
are comparable to the virial equilibrium value prior to disruption.
This leads to the conclusion that the four galactic dSphs with observed
large velocity dispersions must contain dark matter, irrespectively
of whether the systems are in virial equilibrium or are being tidally
disrupted. Very strong tidal interactions of dSphs with the Milky Way,
leading to their disruption, have also been studied by Piatek \& Pryor (1995), 
but they note that their simulations are not relevant to the
Kuhn \& Miller (1989) picture.

This theoretical controversy about the origin of the high velocity dispersions
in dSphs has let to confusion in interpreting
the observations. For example,
Bellazzini et al. (1996) have investigated the
correlation of structural properties of dSphs with their galactocentric
distances (see also Djorgovski \& de Carvalho 1990, Mateo et al. 1993,
Caldwell et al. 1992). Following the model of Kuhn \& Miller (1989),
they conclude that the high velocity dispersions
observed in some dSphs are of tidal origin and that the amount of dark
matter is probably overestimated. In contrast, Irwin \& Hatzidimitriou (1995)
present new observations on the morphology of dSphs. They find a component of
'extra-tidal' stars which they interpret as an evidence for tidal
disruption.  Following Oh et al. (1995), Irwin \& Hatzidimitriou 
conclude that all the dSphs with large M/L contain
significant amounts of dark matter.

Given this situation, it is very important to find more arguments
for or against dark matter in dSphs. This
{\it Letter} will combine recent detailed observations on
the morphology of dSphs with recently published, detailed
numerical calculations. In section 2 it is shown that Draco, Ursa Minor
and Carina cannot contain a dominating dark matter component, if
the interpretation by Irwin \& Hatzidimitriou
of a detected 'extra-tidal' component is correct.
Sextans however might contain significant amounts of dark matter and
an 'extra-tidal' component.  A discussion follows in section 3.

\section{The dark matter content derived from an extra-tidal stellar component.}

Irwin \& Hatzidimitriou (1995) have presented a new determination
of the structural parameters for the four dSphs with exceptionally large M/L
(table 1). With the exception of Sextans, the major axis surface 
brightness profiles can be well fitted by one-component King profiles
with concentration $c=0.5$, 
leading to an accurate determination of the cut-off radii $r_k$.
In addition, beyond $r_k$ an 
excess of so called 'extra-tidal' stars is found, which, according to
Irwin \& Hatzidimitriou (1995)
indicates ongoing tidal stripping. Indeed, the simulations
of Oh et al. (1995) lead to an extended (extra-tidal) component if the
system has a cut-off  radius which is larger than its tidal radius $r_t$.
The cut-off radii of the observed dSphs therefore 
cannot be smaller than their tidal radii ($r_k > r_t$).
The N-body calculations by Oh et al. also show that the dSphs
can only survive the tidal interaction for several Gyrs if $r_k$ is 
smaller than twice the tidal radius ($r_k < 2 \times r_t$). 
This constraint probably applies to the present sample of dSphs,
as it would be very unlikely that $50 \%$ of all galactic satellites
will tidally disrupt and disappear within the next Gyr. In this case, 
the observed cut-off radii provide an estimate
of the tidal radii:

\begin{equation}
0.5 \  r_k < r_t < r_k .
\end{equation}

Sextans is the exception. Its surface brightness profile can be fitted
better by an exponential law or by a King profile with larger concentration:
$c = 1$, resembling typical low-mass
dwarf galaxies (Sandage et al. 1985). Evidence for extra-tidal stars
associated with Sextans has been detected by Gould et al. (1992).
However, the Gould et al. stars, associated with Sextans,
are located within the tidal radius determined by Irwin \&
Hatzidimitriou (1995). They therefore are no longer clearly extra-tidal.
If Sextans is gravitationally bound, we can use again the cut-off
radius of the best fitting King profile as an estimate for $r_t$.

The tidal radius is determined by the orbital parameters of the dSphs. 
Due to uncertainties in the orbital eccentricity e, we 
consider both circular ($e=0$) and
eccentric (e=0.5) orbits. Following Bellazzini et al. (1996),
we assume that the presently observed
galactocentric radius $d$ provides an estimate of the orbital
semi-major axis. The dSphs orbit in the outer galactic
regions, where it is reasonable to
adopt a spherically symmetric galactic potential. We
assume this to be logarithmic, corresponding to a dark matter 
mass distribution of

\begin{equation}
M_G(r) = \frac{v_c^2 r}{G} \approx 1.1 \times 10^{10} \left(
         \frac{r}{kpc} \right) M_{\odot},
\end{equation}

\noindent where G is the gravitational constant and $v_c \approx 220$ km/s is
the constant circular velocity. Within the range of semi-major
axes adopted for the present sample of dSphs, equation 3 should be accurate 
to within a factor of 2 (Da Costa et al. 1991). This uncertainty leads to
errors in determining
the tidal radii of less than a factor 1.3, which is small compared to the
uncertainties in the structural parameters.

Assuming $r_t \approx r_k$ (equation 2), we can rewrite equation 1 as follows

\begin{equation}
\rho_G(d) = \frac{3 M_G(d)}{4 \pi d^3} = \frac{3 \left( \frac{M}{L} \right) L}
     {4 \pi r_k^3} f(e) = \rho_{dSph}.
\end{equation}

\noindent Here f is a function of the orbital eccentricity $e$ and
varies between $f(0) = 1$ and $f(0.5) = 0.42$. 
Only the left hand side of equation 4 ($\rho_G$) depends on the assumed 
galactic mass distribution. The right-hand side ($\rho_{dSph}$) 
is determined by the observed properties of the dSphs (table 1). 
Equation 4 expresses the
fact that the mean mass density of a satellite within its
tidal radius is equal to the mean galactic mass density
within the average orbital radius.

The solid line in Figure 1 shows the average galactic mass density $\rho_G(d)$,
calculated using equations 3 and 4. The open circles show $\rho_{dSph}$
for the present sample of dSphs, assuming $M/L=1$. The filled circles show 
$\rho_{dSph}$ assuming a $M/L$ value derived dynamically from
the observed velocity dispersions (table 1).

If the extended stellar component in the dSphs is extra-tidal, the cut-off
radii should provide a good estimate of the tidal radii (equation 2).
The data points then should lie on the solid line.
Even with the large error bars, which are due to the
large observational uncertainties in the orbital
and structural parameters, figure 1 clearly demonstrates that in this case
Carina, Draco and Ursa Minor could not be dark matter dominated
(see also Moore 1996).
If their masses were as high as suggested from their velocity dispersions,
their internal densities (filled circles) would be much larger than expected
for a tidally limitated satellite.
No 'extra-tidal' stars should be observed as $r_t \gg r_k$.

Sextans has a cut-off radius which is
consistent with the assumption of a large $M/L$.
The good agreement of its kinematically derived average
density (filled circle) with the expected value provides strong support for
the conclusion that this satellite is dominated by dark matter.
If Sextans is a gravitationally bound satellite, then it must contain
a dark matter halo as otherwise its
tidal radius would be a factor of five smaller than its
cut-off radius. The numerical simulations by Oh et al. (1995)
then would indicate (see equation 2) that Sextans should completely disrupt 
within the next $10^8$ yrs.
It is very unlikely that we observe this satellite in such a short-lived
evolutionary state.  In addition, in this case Sextans should have higher
ellipticities than observed.

\section{Discussion}

The solid line in figure 1 subdivides the diagram into two regions. Only
those satellites which lie below this line should be tidally affected.
The existence of 'extra-tidal' stars in Carina, Draco and Ursa Minor therefore
would demonstrate that these dSphs cannot contain significant amounts of dark matter.
Note that this conclusion is independent of the accuracy
by which the velocity dispersion is determined. Any value of
$M/L$ which is larger than 3 would lead to such high
internal densities that the cut-off radii could not be of order the
tidal radii (Pryor 1996).
If Sextans is not in the process of complete tidal disruption it must 
contain  a dominating dark matter component.
Such a difference in the dark matter content of the galactic satellites
could result from different formation histories. It is intriguing that
Carina, Draco and Ursa Minor lie in the Magellanic plane and might originate
from a tidally disrupted progenitor satellite, as suggested by Lynden-Bell (1982).
In this case, no significant dark matter component should indeed be found.
In contrast, Sextans cannot be associated with the Magellanic plane 
(Lynden-Bell 1993). This object might represent a typical low-mass,
low-surface density dwarf galaxy which was captured and still contains
at least the inner parts of its original dark matter halo.

Up to now we have assumed that there exists an 'extra-tidal' stellar 
component. However, the previous results seem to be in
conflict with the observed high velocity dispersions in Carina, Ursa Minor
and Sextans. The computations by Oh et al. (1995) demonstrate that,
without dark matter, the velocity dispersion could not be as high as
observed, even if tidal heating is taken into account. 
High dispersions might be measured in systems which are
contaminated by close binary stars (Suntzeff et al.
1993). However, Armandroff et al. (1995) have shown that this effect can
be ruled out in the case of the dwarf spheroidals. The most reasonable
explanation seems to be that all four dwarf spheroidals contain 
massive dark matter halos. With the exception of Sextans,
the extended stellar component then should not be classified as 'extra-tidal', 
because it actually lies deeply 
embedded in the potential well of the dwarf spheroidal's dark matter halos.
How such an extended component could arise is unknown. 
One also might wonder about
the puzzling coincidence that the observed cut-off radii of these dwarf
spheroidals are very similar to the expected tidal radii, assuming no
dark matter component.

More detailed observations and accurate velocity measurements
of the extended stellar component
are required in order to clarify whether there really exists a 
break to a shallower slope
in the density profiles of the dwarf spheroidals and whether
the outermost stars are tidally stripped.  Such observations
will also clarify the important question whether the dwarf spheroidals
contain a dominant dark matter component or not.

\acknowledgments

I would like to thank the referee, Dr. Pryor, for important suggestions.

\clearpage

\begin{table*}
\begin{center}
\begin{tabular} {lcccc}
Name & $L (L_{\odot}$) & $d (kpc)$ & $r_k$ (pc) & $\frac{M}{L}
      \left( \frac{M_{\odot}}{L_{\odot}} \right) $ \\
\tableline
Carina &$(2.4 \pm 1.0) \times 10^5$ &$85 \pm 5$ &$581 \pm 86$ &$59 \pm 47$\\
Draco  &$(1.8 \pm 0.8) \times 10^5$ &$72 \pm 3$ &$498 \pm 47$ &$245 \pm 155$\\
Ursa Minor &$(2.0 \pm 0.9) \times 10^5$ &$64 \pm 5$ &$628 \pm 74$ &$95 \pm 43$\\
Sextans &$(4.1 \pm 1.9) \times 10^5$ &$83 \pm 9$ &$3102 \pm 1028$ &$107 \pm 72$\\
\end{tabular}
\end{center}

\tablenum{1}
\caption{The sample of galactic dSphs with large $M/L$. The physical
         properties are adopted from Irwin \& Hatzidimitrou (1995),
         with $L$ the total luminosity, $d$ the galactocentric distance, 
         $r_k$ the cut-off radius and $M/L$ the dynamically derived
         mass-to-light ratio.}

\end{table*}

\clearpage

\newpage
\begin{figure}
{\bf Figure 1} The solid line shows the average galactic mass density 
$\rho_G$ (equation 3) as a function of galactocentric distance $d$. 
The open circles show the average internal densities
$\rho_{dSph}$ (equation 3) of the dSphs for $M/L = 1$. The filled circles
show $\rho_{dSph}$, adopting the kinematically derived $M/L$ (table 1).
Error bars show the effect of the uncertainties in the galactocentric 
distance, orbital eccentricity, tidal radius and luminosity.
In sequence of increasing $d$, the pairs of open and filled
circles correspond to Ursa Minor, Draco, Sextans and Carina.
dispersion $\sigma_r(r)$ and the dashed line shows the tangential velocity
dispersion $\sigma_t(r)$.
\input epsf
\centering
\leavevmode
\epsfxsize=1.0
\columnwidth
\epsfbox{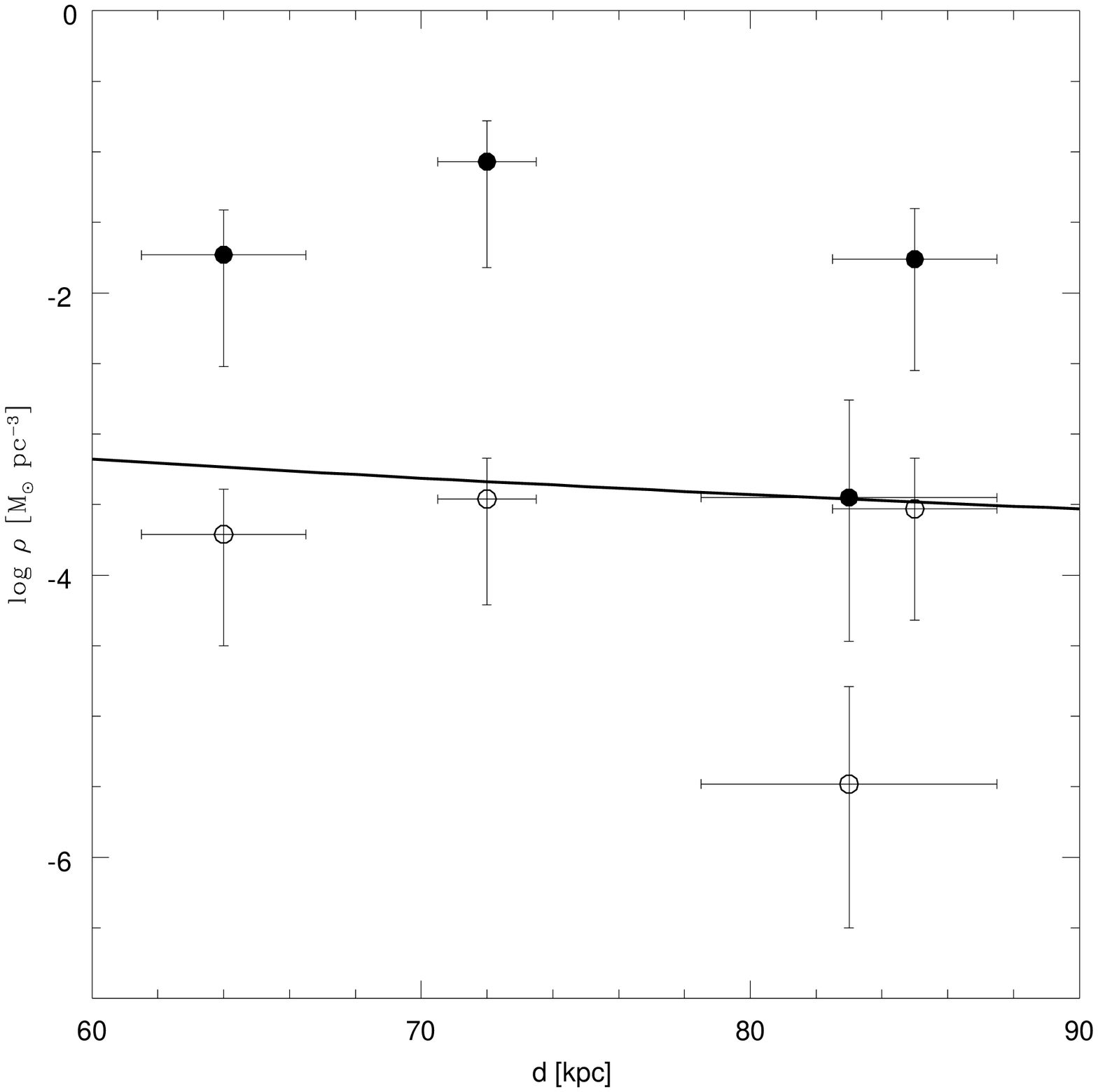}
\end{figure}

\end{document}